\documentclass[conference]{IEEEtran}
\IEEEoverridecommandlockouts
\usepackage{cite}
\usepackage{amsmath,amssymb,amsfonts}
\usepackage{algorithmic}
\usepackage{graphicx}
\usepackage{textcomp}
\usepackage{xcolor}
\def\BibTeX{{\rm B\kern-.05em{\sc i\kern-.025em b}\kern-.08em
    T\kern-.1667em\lower.7ex\hbox{E}\kern-.125emX}}

\usepackage{graphicx}
\usepackage{multirow,array}
\usepackage{subcaption}
\usepackage{multicol}  
\usepackage{amsmath}
\usepackage{hyperref}
\usepackage{todonotes}
\usepackage[utf8]{inputenc} 
\usepackage[T1]{fontenc}    
\usepackage{hyperref}       
\usepackage{url}            
\usepackage{booktabs}       
\usepackage{amsfonts}       
\usepackage{nicefrac}       
\usepackage{microtype}      
\usepackage{caption}
\usepackage{multirow}
\usepackage{subcaption}
\usepackage[ruled,vlined]{algorithm2e}
\usepackage{enumitem}
\usepackage{bm}
\hypersetup{
    colorlinks=true,
    linkcolor=blue,
    filecolor=magenta,      
    urlcolor=cyan,
}
\IEEEoverridecommandlockouts

\usepackage{graphics}

\renewcommand{\SS}{\mathcal{S}}

\newcommand{\eat}[1]{}

\begin{document}
\author
{
 Eugene Vinitsky\IEEEauthorrefmark{1},
 Nathan Lichtle\IEEEauthorrefmark{5},
  Kanaad Parvate\IEEEauthorrefmark{2},
    Alexandre M Bayen\IEEEauthorrefmark{2}\IEEEauthorrefmark{3}
\thanks{\noindent Corresponding author: Eugene Vinitsky (evinitsky@berkeley.edu)}
\thanks{Email addresses:\{kanaad, evinitsky, bayen\} @berkeley.edu, nathan.lichtle@ens-paris-saclay.fr} \\
    \IEEEauthorblockA{\IEEEauthorrefmark{1}UC Berkeley, Department of Mechanical Engineering} 
    \IEEEauthorblockA{\IEEEauthorrefmark{2}UC Berkeley, Electrical Engineering and Computer Science} 
    \IEEEauthorblockA{\IEEEauthorrefmark{3}UC Berkeley, Institute for Transportation Studies}
    \IEEEauthorblockA{\IEEEauthorrefmark{4}Paris-Saclay University, ENS Paris-Saclay, Department of Computer Science}
}

\title{Optimizing Mixed Autonomy Traffic Flow With Decentralized Autonomous Vehicles and Multi-Agent RL}


\maketitle

\begin{abstract}
We study the ability of autonomous vehicles to improve the throughput of a bottleneck using a fully decentralized control scheme in a mixed autonomy setting. We consider the problem of improving the throughput of a scaled model of the San Francisco-Oakland Bay Bridge: a two-stage bottleneck where four lanes reduce to two and then reduce to one. Although there is extensive work examining variants of bottleneck control in a centralized setting, there is less study of the challenging multi-agent setting where the large number of interacting AVs leads to significant optimization difficulties for reinforcement learning methods. We apply multi-agent reinforcement algorithms to this problem and demonstrate that significant improvements in bottleneck throughput, from 20\% at a 5\% penetration rate to 33\% at a 40\% penetration rate, can be achieved. We compare our results to a hand-designed feedback controller and demonstrate that our results sharply outperform the feedback controller despite extensive tuning. Additionally, we demonstrate that the RL-based controllers adopt a robust strategy that works across penetration rates whereas the feedback controllers degrade immediately upon penetration rate variation. We investigate the feasibility of both action and observation decentralization and demonstrate that effective strategies are possible using purely local sensing. Finally, we open-source our code at \url{https://github.com/eugenevinitsky/decentralized_bottlenecks}.
\end{abstract}

\section{Introduction}

The last few years have seen the widespread, successful deployment of human-in-the-loop autonomous vehicle (AV) systems. Every major vehicle manufacturer offers some variant of a camera-based level-two system (autonomous distance and lane keeping), paving the way for a gradual transition from primarily human-driven transportation systems to a mixture of automated and human driven vehicles, a regime referred to as \emph{mixed autonomy traffic}. Even at low penetration rates of AVs, this period offers an exciting opportunity to reshape the efficiency of transportation systems by using the AVs as mobile controllers. AVs have fast reaction times, superior sensing capabilities, and can be programmed to optimize socially desirable objectives like improved throughput and lowered energy consumption. In this work we will demonstrate how level-2 AVs, equipped with standard sensors like cameras and radars, can be used to improve the throughput of a simplified model of the San Francisco-Oakland Bay Bridge and similar bottleneck structures.



We will focus on using AVs to improve the throughput of a lane reduction, a road architecture where the number of lanes suddenly decreases. We will refer to successive lane reductions as a \emph{traffic bottleneck}. Bottlenecks are believed to cause a phenomenon known as \emph{capacity drop}~\cite{hall1991freeway, chung2007relation} where the inflow-outflow relationship at the bottleneck is initially linear but above some critical inflow value experiences a hysteric transition where the outflow suddenly and sharply drops (see Fig.~\ref{fig:capacity} for an example). The imbalance between inflow and outflow leads to congestion and a reduction in the throughput of the bottleneck. 

To avoid this reduction, it is necessary to restrict the inflow so that it never exceeds the critical value above which capacity drop occurs. One approach to tackling this is to introduce traffic lights into the network that meter/restrict the inflow~\cite{papageorgiou1991alinea} but this would require the installation of additional infrastructure. Instead, autonomous vehicles can be used as mobile metering infrastructure, essentially distributed traffic lights, that intelligently select when to meter and when to let the flow continue without restriction.

While there is work characterizing bottleneck control using vehicle-based control, it usually operates in the centralized regime where a single controller outputs commands to all the AVs in the system either via variable speed limits~\cite{smulders1990control, wu2018differential, lu2010new, iordanidou2014feedback} or centrally coordinated platoons~\cite{vcivcic2019coordinating}. Here we consider the challenging multi-agent problem where each AV operates in a fully decentralized fashion and control is applied at the level of individual vehicle accelerations. The AVs can still coordinate but only implicitly: they can only use common knowledge to decide which AV should go next. While decentralization adds additional difficulty in controller design, the resultant controllers should be realizable using existing cruise control technology and can consequently be implemented without relying on any improvements in vehicle-to-vehicle communication technology.

We investigate the potential impact of decentralized AV control on bottleneck throughput by studying a scaled-down version of the post-tollbooth section (see Fig.~\ref{fig:baybridge}) of the San Francisco-Oakland Bay Bridge. In our scaled version, four lanes reduce to two which then reduce down to one lane (as opposed to 15 to 8 to 5 in the bridge). While the lane numbers differ, the overall road architecture is quite similar as each vehicle goes through two merges. To design the controllers, we will use multi-agent reinforcement learning (MARL). Even at reduced scale, this problem is a difficult MARL challenge as it incorporates:
\begin{itemize}
    \item a large number of agents, varying between 20-200 depending on the penetration rate.
    \item delayed reward structure. A given vehicle's impact on outflow isn't experienced until many seconds later.
    \item challenging credit assignment. The outflow is a global signal and it is difficult to disambiguate whose action led to the improved outcome.
\end{itemize}
This work tackles this challenging MARL problem and provides some initial characterization of the performance of decentralized control in these settings. The main contributions of this work are:
\begin{enumerate}
\item We introduce a challenging new benchmark in multi-agent reinforcement learning.
\item We demonstrate that appropriately chosen multi-agent RL algorithms can be used to design decentralized control policies for maximizing bottleneck throughput.
\item We show that effective control can be performed in the fully local sensing setting where vehicles do not have access to any macroscopic observations.
\item We demonstrate and formalize a challenging problem in open transportation networks where the Nash equilibrium can deviate from the social equilibrium. We introduce a simple trick to make the two equilibria align.
\item We design decentralized feedback control policies and show that, despite extensive tuning, the RL policies sharply outperform our feedback baseline. Additionally, the RL approach is able to equal the performance of a traffic-light baseline.
\item We demonstrate that the resultant control policies can be made robust to variations in the penetration rate.
\end{enumerate}

The rest of the article is organized as follows. 
Section~\ref{sec:background} provides an introduction to deep RL, off-policy RL methods, car following models, and \emph{Flow}, the traffic control library used in this work. It also introduces our feedback control baseline and traffic light baseline. 
Section~\ref{sec:experiments} formulates the capacity drop diagrams of our bottleneck and explains the state and action spaces for all the controllers studied herein. Section~\ref{sec:results} provides a discussion of the results. Finally Section~\ref{sec:conclusions} summarizes our work and provides a discussion of possible future research directions. 

\section{Background}
\label{sec:background}

\subsection{Reinforcement Learning}
In this section, we discuss the notation and describe in brief the key ideas used in reinforcement learning. Reinforcement learning focuses on maximization of the discounted reward of a \textit{Markov decision process}  (MDP)~\cite{bellman1957markovian} or partially observed Markov decision process (POMDP) in which the agent has restricted access to the true world state.
The system described in this article solves tasks which conform to the standard structure of a finite-horizon discounted multi-agent POMDP, defined by the tuple
$(\SS_0, \mathcal{A}_0, \mathcal{O}_0, r_0, \rho_0, \gamma_0, T_0) \times \dots \times (\SS_n, \mathcal{A}_n, \mathcal{O}_n, r_n, \rho_n, \gamma_n, T_n) \times \times P \times \mathcal{Z}$, where $\SS_i$ is a (possibly infinite) set of states for agent $i$, $\mathcal{A}_i$ is a set of actions for agent $i$, $\mathcal{Z}: (\SS_0 \times \mathcal{A}_0) \times \dots \times (\SS_n \times \mathcal{A}_n) \to (\mathcal{O}_0, \dots, \mathcal{O}_n)$ is a function describing how the world state is mapped into the observations of the POMDP, $P: (\SS_0 \times \mathcal{A}_0 \times \SS_0) \times \dots \times (\SS_n \times \mathcal{A}_n \times \SS_n) \to \mathbb{R}_{\geq0}$ is the transition probability distribution for moving from one set of agent states $s$ to the next set of states $s'$ given the set of actions $(a_0, \dots, a_n)$,  $r_i : (\SS_0 \times \mathcal{A}_0) \times \dots \times (\SS_n \times \mathcal{A}_n) \to \mathbb{R}$ is the reward function for agent $i$, $\rho_i: \SS_i \to \mathbb{R}_{\geq 0}$ is the initial state distribution for agent $i$, $\gamma_i \in (0, 1]$ is the discount factor for agent $i$, and $T_i$ is the horizon for agent $i$. 

RL studies the problem of how an agent can learn to take actions in its environment to maximize its cumulative discounted reward. Specifically it tries to optimize $J^\pi = \mathbb{E}_{\rho_0, \ p(s_{t+1}|s_t, a_t)}\left[\sum_{t=0}^T \gamma^t r_t  \mid \pi(a_t | s_t)\right]$ where $r_t$ is the reward at time $t$ and the expectation is over the start state distribution, the probabilistic dynamics, and the probabilistic controller $\pi$. Note that we have temporarily dropped the dependence on agent index for the purpose of clarity. The goal in RL is to use the observed data from the MDP to compute the controller $\pi: \SS \to \mathcal{A}$, mapping states to actions, that maximizes $J^\pi$. It is increasingly popular to parametrize the controller as a neural network. We will denote the parameters of this controller, in this case the neural network weights, by $\theta$ and the controller by $\pi_\theta$. A neural net consists of a stacked set of affine linear transforms and non-linearities that the input is alternately passed through. The presence of multiple stacked layers is the origin of the term "deep" reinforcement learning. In this work we will use a shared Multi-Layer Perceptron (MLP); each agent will use the exact same controller. Details of the architecture are provided in Sec.~\ref{sec:exp_details}.

\subsection{Off-Policy Reinforcement Learning}
\label{sec: TD3}
Here we briefly introduce off-policy reinforcement learning methods in an attempt to clarify some of the difficulties of using single-agent algorithms in multi-agent settings. For a more thorough discussion of the underlying algorithms see~\cite{fujimoto2018addressing, lillicrap2015continuous} and for the particular challenges of multi-agent off-policy algorithms see~\cite{lowe2017multi}.

Off-policy methods focus on using a buffer of data sampled from the environment to construct the policy. While they can suffer from instability relative to policy gradient methods~\cite{achiam2019towards}, they tend to be more sample efficient and can often be effectively run on a single CPU. The basic idea is to periodically sample data from the buffer and compute an estimate of the Bellman error
\begin{equation*}
    \mathcal{L} = \frac{1}{N}\sum_{i=1}^N \left(Q(s^i_t, a^i_t) - r(s^i_t, a^i_t) - \gamma \, \underset{a}{\text{argmax}}\,Q(s^i_{t+1}, a) \right)^2
\end{equation*}
where $i$ indexes a sample from the batch, $\gamma$ is the discount factor, and Q is the Q-function
\begin{equation*}
    Q^\pi(s_t, a_t) = \mathbb{E}_\pi\left[\sum_{i=t}^T \gamma^{i-t}r_t(s_t, a_t) |s_t, a_t\right]
\end{equation*}
i.e. the expected cumulative discounted reward of taking action $a_t$ and thereafter following policy $\pi$ (we will use the terms policy and controller interchangeably in this section). The sum $r(s^i_t, a^i_t) + \gamma \, \underset{a}{\text{argmax}}\,Q(s^i_{t+1}, a)$ is referred to as the \emph{target}. The algorithms then perform gradient descent on the loss $\mathcal{L}$ to learn an approximation of the Q-function.

In this work we use Twin-Delayed Deep Deterministic Policy Gradient (TD3)~\cite{fujimoto2018addressing} a variant of Deep Deterministic Policy Gradient (DDPG)~\cite{lillicrap2015continuous}. DDPG simultaneously learns a Q-function for estimating the values of states and a policy that selects actions that maximize the Q-function. Both policy $\mu$ and Q-function are learned simultaneously: the Q-function is learnt by minimizing the Bellman error over a batch of data using gradient descent and the policy $\mu$ is learnt by performing gradient ascent on the action component of the Q-function. 

TD3 creates an empirically stabler version of DDPG by adding three simple tricks:
\begin{itemize}
    \item The target is estimated using two Q-functions instead of one and taking the minimum of the two.
    \item The policy network is updated significantly less often than the value network.
    \item A small amount of noise is added to the action when estimating the value of the Q-function. This is based on the assumption that similar actions should have similar Q-values.
\end{itemize}
For more details, please refer to~\cite{fujimoto2018addressing}. 

It is important to note that TD3 is a single agent algorithm and that in multi-agent settings, there is additional instability induced by the changing policies of the other agents in the environment. Essentially, because the other agents in the environment have also changed, samples inside the buffer are \emph{stale}, they no longer correctly represent either the rewards that would be received for taking an action in a given state nor is the subsequent state after taking that action correct. Algorithms that fail to address this issue and simply perform Q-learning while ignoring it are referred to as \emph{Independent Learners}.
This challenge can be addressed by algorithms like Multi-agent DDPG~\cite{lowe2017multi} which use a Q-function that sees the states and actions of all active agents. 

In this work we simply use \emph{Independent Learners} with a shared policy: all of our agents use the same neural network to compute their actions. Surprisingly, we find this to be effective despite the issue of \emph{stale} buffer samples discussed above.

\subsection{Car Following Models}
\label{sec:car-following}
For our model of the driving dynamics, we use the default car following model and lane changing models in SUMO. We use SUMO 1.1.0 which has the Krauss car following model~\cite{krauss1998microscopic}. The Krauss car following model is quite simple: namely the ego vehicle drives as fast as possible (subject to a maximum speed) while keeping a distance such that if the lead vehicle brakes as hard as possible, the ego vehicle is able to safely stop in time. For the parameters of the model, we use the default values in the aforementioned SUMO version. The lane changing model is also the default model described in~\cite{erdmann2015sumo}.
\subsection{Flow}
We run our experiments in \texttt{Flow}~\cite{wu2017flow}, a library that provides an interface between the traffic microsimulators SUMO~\cite{SUMO2018} and AIMSUN~\cite{barcelo2005dynamic}, and RLlib~\cite{liang2018rllib}, a distributed reinforcement learning library. \texttt{Flow} enables users to create new traffic networks via a python interface, introduce autonomous controllers into the networks, and then train the controllers on many-CPU machines on the cloud via AWS EC2. To make it easier to reproduce our experiments or to try and improve on our results, our fork of \texttt{Flow}, scripts for running our experiments, reproducing our results, and tutorials can be found at \url{https://github.com/eugenevinitsky/decentralized_bottlenecks}.


\subsection{Feedback Control and ALINEA}
\label{sec:feedback_control}
As our baseline for the performance of our RL controllers, we implement the traffic light controller from~\cite{papageorgiou1991alinea} (referred to here as ALINEA) and additionally design a decentralized variant of ALINEA that can be performed using AVs. The basic idea underlying ALINEA is to select an optimal bottleneck vehicle density and then perform feedback control around that optimal value using the ratio of red-time to green-time of the traffic light as the control parameter. We use the particular scheme outlined in~\cite{spiliopoulou2010toll} with some slight modifications.

Instead of operating around density, we feedback around a desired number of vehicles in the bottleneck which we denote as $n_\text{crit}$, a hyperparameter that we will empirically determine for our network. We update the desired inflow $q$ as 
\begin{gather*}
    \tilde{q}_{k+1} = q_k + K(n_\text{crit} - \hat{n}) \\
    q_{k+1} = \text{max}(q_\text{min}, \text{min}(\tilde{q}_{k+1}, q_\text{max}))
\end{gather*}
where K is the gain of the proportional feedback controller, $\hat{n}$ is the average number of vehicles in the bottleneck over the last T seconds, and $q_\text{min}$ and $q_\text{max}$ are minima and maxima of $q$ to prevent issues with wind-up. We set $T=25, \, q_\text{min}=200, \, q_\text{max}=14400,$ and perform hyperparameter searches over $K, n_\text{crit}, q_0$. This desired inflow is then converted into a red-green cycle time via
\begin{equation*}
    c_k = r + g = \frac{7200 * L }{q_k}
\end{equation*}
where $r$ is the red time, $g$ is a fixed green time, and $L$ is the maximum number of lanes. In this work, we set $g$ to $4$ which we empirically determined to be the amount of time needed to let two vehicles pass into the bottleneck. We perform this feedback update every $30$ seconds. Finally, we initialize each of the traffic lights to have a cycle that is offset from each other by 2 seconds to prevent the traffic lights from being completely in sync. Further details are provided in the code.

Fortunately, we can apply the exact same strategy using autonomous vehicles instead of traffic lights where $c_k$ is now the amount of time that an AV will wait at the bottleneck entrance before entering. However, the hyper-parameters will differ sharply as a function of the penetration rate. For a given penetration rate percentage, $p$, the expected length of the human platoon behind a given AV will be $\frac{1}{p} - 1$. Whereas in the traffic light case we can set arbitrary inflows, here every time an AV goes, $\frac{1}{p} - 1$ vehicles will follow it on average. As a consequence, to avoid congestion at lower penetration rates, the inflow needs to be a good deal lower. As we will discuss in Section \ref{sec:results}, this leads to the decentralized control scheme under-performing traffic-light based control. For the exact hyper-parameters swept, see the appendix.

\section{Experiments}
\label{sec:experiments}
\subsection{Experiment setup}
\begin{figure}
\centering
\includegraphics[width=0.5\textwidth]{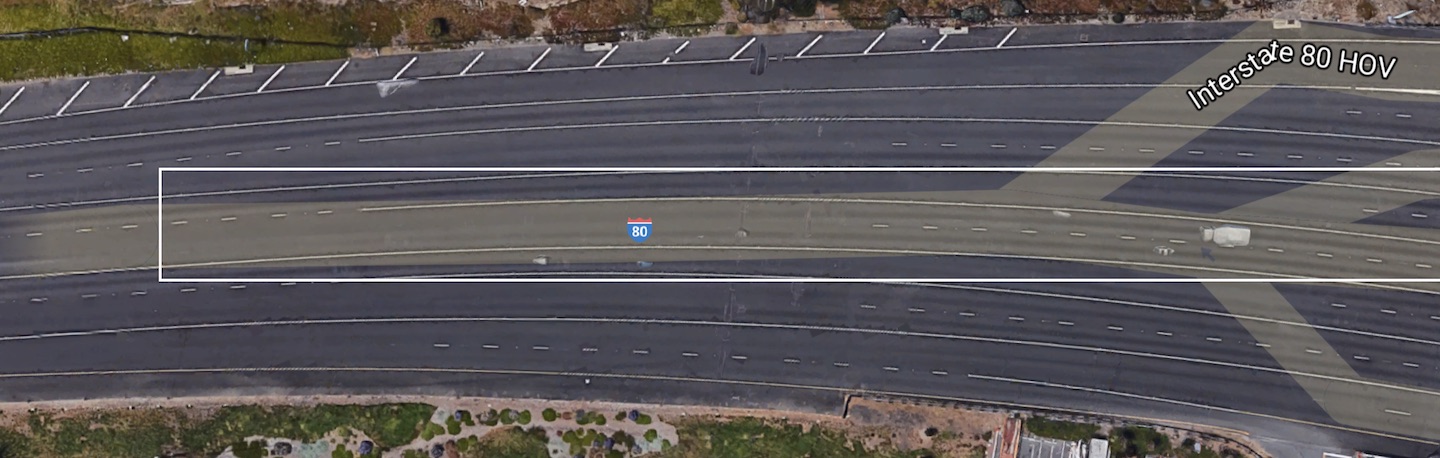}
\caption{Bay bridge merge. Equivalent subsection that we study in this work is highlighted in a white square. Traffic travels from right to left.}
\label{fig:baybridge}
\end{figure}

\begin{figure}
\includegraphics[width=0.5\textwidth]{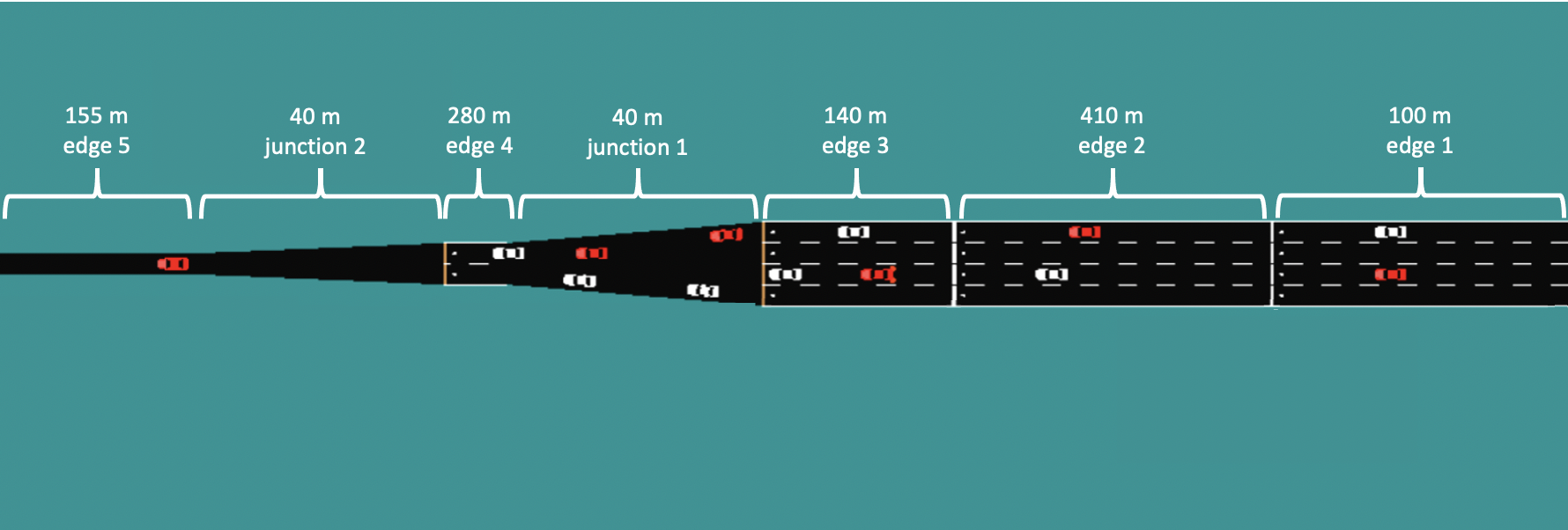}
\caption{Long entering segment followed by two zipper merges, a long segment, and then another zipper merge. Red cars are automated, human drivers are in white. Scale is severely distorted to make visible relevant merge sections.}
\label{fig:segments}
\end{figure}

\begin{figure}
\centering
\includegraphics[width=0.44\textwidth]{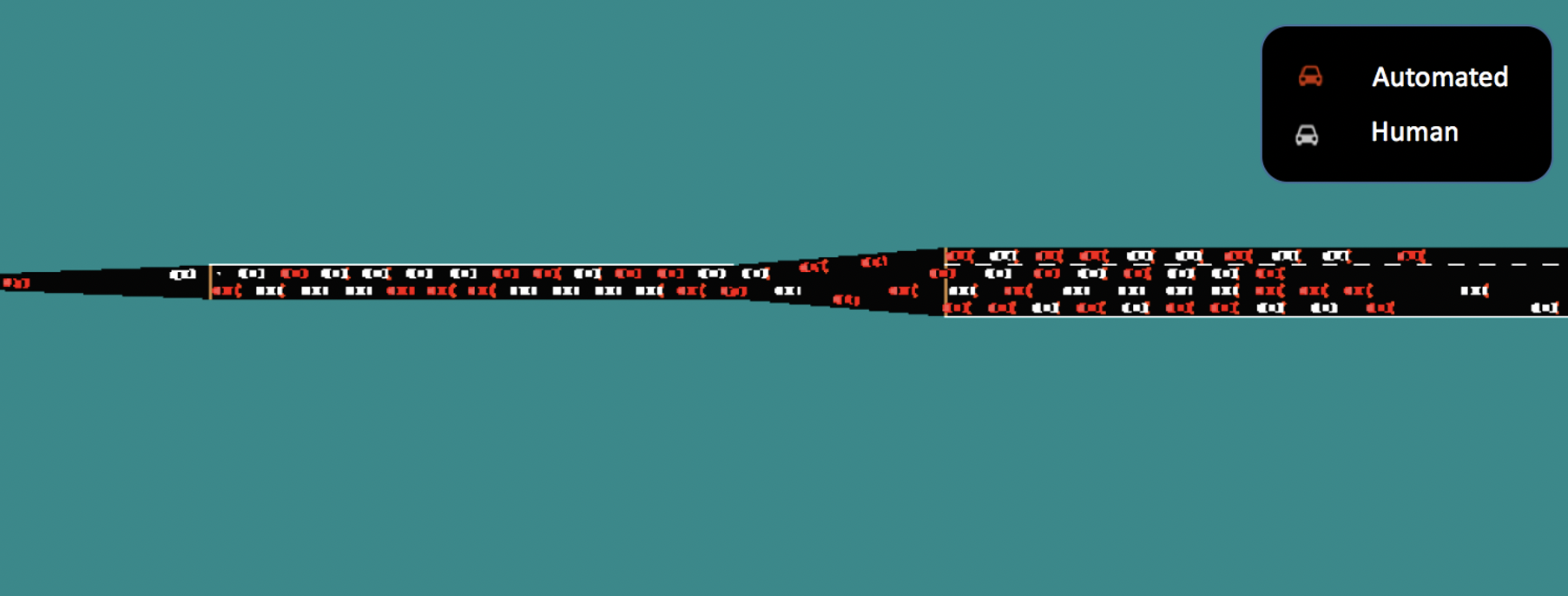}
\caption{Congestion forming in the bottleneck. Congestion starts at the left and propagates right.}
\label{fig:bottleneck_congestion}
\end{figure}

We attempt to improve the outflow of the bottleneck depicted in Fig. \ref{fig:segments}, in which a long straight segment is followed by two zipper merges sending four lanes to two, and then another zipper merge sending two lanes to one. This is a simplified model of the post-ramp meter bottleneck on the Oakland-San Francisco Bay Bridge depicted in Fig.\,\ref{fig:baybridge}.
Once congestion forms, as in Fig. \ref{fig:bottleneck_congestion}, the congestion is does not dissipate due to lower outflow than inflow and begins to extend upstream.

An important point to note is that in this work lane changing is disabled for all the vehicles in this system. As we discuss in Sec.~\ref{sec:ablations}, this enables higher throughput but would require the imposition of new road-rules at the bottleneck. Fortunately, this would only require painting some new lines that restrict lane changing which should be relatively cheap.


\subsection{Capacity diagrams}
\label{sec:capacity}
\begin{figure}
\centering
\includegraphics[width=0.5\textwidth]{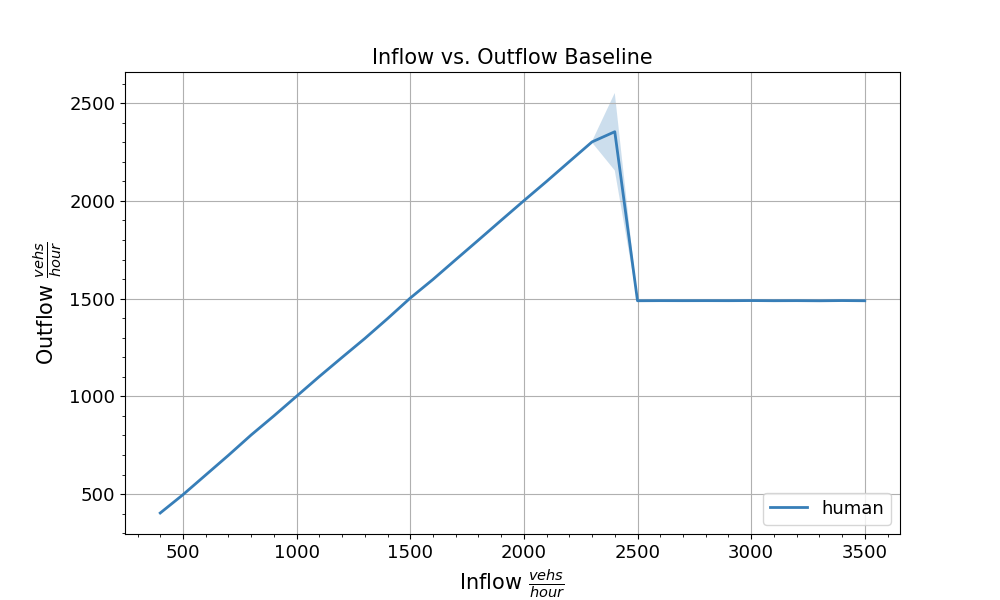}
\caption{Inflow vs. outflow for the uncontrolled bottleneck. The solid line represents the average over 20 runs at each inflow value and the darker transparent section is the one standard deviation from the mean.}
\label{fig:capacity}
\end{figure}
Fig.\,\ref{fig:capacity} presents the inflow-outflow relationship of the uncontrolled bottleneck model. To compute this, we swept over inflows from 400 to 3500 vehicles per hour in steps of 100, ran 20 runs for each inflow value, and took the outflow as the average outflow over the last 500 seconds. Fig.\,\ref{fig:capacity} presents the average value and 1 std-deviation from the average across these 20 runs. Below an inflow of 2300 vehicles per hour congestion does not occur; above 2500 vehicles per hour congestion will form with high certainty. A key point is that once the congestion forms at these high inflows, at values upwards of 2500 vehicles per hour, it does not dissolve unless inflow is reduced for a long period of time. Identical inflow-outflow behavior is observed when lane changing is enabled so a similar graph with lane changing is omitted.


\subsection{Partially Observed Markov Decision Process Structure}
\label{sec:mdp_construction}
Here we outline the definition of the action space, observation function, reward function, and transition dynamics of the POMDP that is used in our controllers. We distinguish three cases that depend on the type of sensing infrastructure that will be available to the controller. The central concern is that the observations must give some way of identifying the state of the bottleneck (speed and density) to allow the AVs to intelligently regulate the inflow. Without some estimate of bottleneck state, the AVs must be extremely conservative to ensure that the bottleneck does not enter congestion. The estimation of bottleneck state can be done explicitly by acquiring the bottleneck state with loop detectors/overhead cameras or implicitly, by observing the behavior of vehicles around the bottleneck and inferring what the state of the bottleneck must be.

\begin{figure}
\centering
\includegraphics[width=0.48\textwidth]{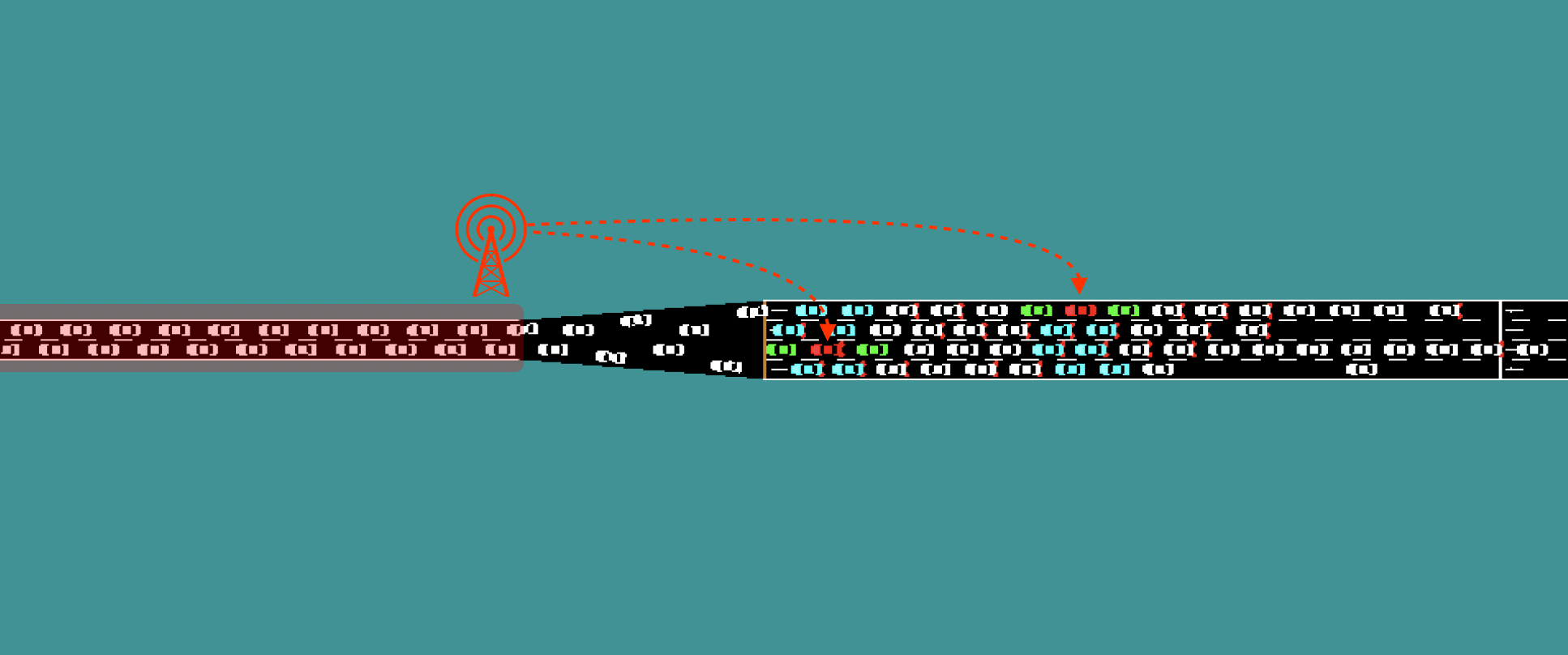}
\caption{Diagram representing the different state spaces. The red vehicles can see the green vehicles in the minimal state space, and the blue and green vehicles in the radar state space. In the minimal and aggregate state spaces, the red vehicle also has access to information about the vehicle count in the bottleneck which we represent as a tower communicating information about the highlighted red segment. In addition to the states indicated here, the aggregate state space also contains the average speeds of edges 3, 4, and 5 (see Fig.~\ref{fig:segments}).}
\label{fig:state_spaces}
\end{figure}

Figure.~\ref{fig:state_spaces} provides a rough overview of our considered state spaces. We call the first state set the \emph{radar state} as the required states would be readily available via on-board radar, cameras, and GPS. This is the state space that would be most easily implemented using existing technology on an autonomous vehicle and does not use any macroscopic information. In Fig.~\ref{fig:state_spaces}, \emph{radar state} would consist of the distances and speeds of the blue and green vehicles. We also provide a small state space we call the \emph{minimal} state space which essentially consists of the speed and distance of the green car in Fig.~\ref{fig:state_spaces} as well as the vehicle counts in the bottleneck. This is a small state space intended for fast learning. We also investigate an \emph{aggregate} state that provides macroscopic data about the bottleneck and can be added on to any existing state space. The aggregate state consists of the number of vehicles in the bottleneck and the average speeds of vehicles in edges 3, 4 and 5 (see Fig.~\ref{fig:segments} for the numbering). Theses states would be available given appropriate loop sensing infrastructure or a sufficient number of overhead cameras distributed throughout the bottleneck. Finally, we note that every state space contains the ego vehicle speed, its GPS position on the network, and a counter indicating how long the vehicle has stopped. This counter is used to enable the controller to track how long it has waited to enter the bottleneck. 

From these three potential sets of states we form three combined state spaces that we study: radar + aggregate, minimal + aggregate, and minimal alone. Each of these represents a different set of assumptions on what sensing technology will be available ranging from full decentralization (radar alone) to having access to macroscopic information (minimal, aggregate). We characterize the relative performance of these different state spaces in Sec.~\ref{sec:sensing_exps}.

The action space is simply a 1-dimensional acceleration. 
While we could include lane changes as a possible action, we leave this to future work. To prevent the vehicles from forming unusual patterns at the entrance of the bottleneck, control is only applied on edge 3 (edges numbered according to Fig.~\ref{fig:segments}). However, states and rewards are received at every time-step and consequently actions are computed at each time-step: we simply ignore the controller output and use the output of the car following model unless we are on edge 3.

We are trying to optimize throughput, so for our reward function we simply use the number of vehicles that have exited in the previous time-step as a reward
\begin{equation*}
    r_t(s_t, a_t) = n_t / N
\end{equation*}
where $n_t$ is the number of vehicles that have exited in that time-step and $N$ is a normalization term that was used to keep the cumulative reward at reasonable values. We use $N=50$ in this work.
Since the outflow is exactly the quantity we are trying to optimize, optimizing our global reward function should result in the desired improvement. This is a global reward function that is shared by every agent in the network. 

However, we note a few challenges that make this a difficult reward function to optimize. First, the reward is global which causes difficulties in credit assignment. Namely, it is not clear which vehicle's action contributed to the reward at any given time-step. Secondly, there is a large gap between when an action is taken and when the reward is received for that action. That is, a vehicle choosing to enter the bottleneck does not receive any reward directly attributable to that decision for upwards of 20 steps. Finally, the bottleneck being fully congested is likely a local minimum that is hard to escape. Once congestion has onset, it cannot be removed without a temporary period where the inflow into the bottleneck is reduced. However, a single vehicle choosing to not enter the bottleneck would have negligible effect on the inflow, making it difficult for vehicles to learn that decongestion is even possible.

For more details on the POMDP, see Appendix Sec.~\ref{sec:detailed_mdp}

\subsection{Divergence between Nash Equilibrium and Social Optimum}
\label{sec:nash_social}
Here we provide a simple illustrative example of how, despite having a single, global reward function, open networks can lead to non-cooperative behavior. In the case where every vehicle receives the same reward at every time-step, it is simple to see that the Nash Equilibrium will be the same as the social optimum. However, 
vehicles sharing the same reward function but optimizing over different horizons can cause a divergence between the two equilibria. The key intuition is that although all of the vehicles are trying to optimize the same quantity, they only receive rewards while they are in the system as their trajectory terminates once they go through the exit. This creates a perverse incentive to remain in the system for longer than is socially desirable, leading to a divergence from the social optimum.
Consider the following simplified, single-step variant of the bottleneck in which there are simply two vehicles. The problem has the following reward structure before we introduce the open-endedness:
\begin{itemize}
    \item If both vehicles go, congestion occurs and they receive a reward of 1.
    \item If one vehicle goes and the other doesn't, no congestion occurs. Both vehicles receive a reward of 2.
    \item If both vehicles do not go, no outflow occurs and they receive a reward of 0.
\end{itemize}
\begin{table}
\centering
    \setlength{\extrarowheight}{4pt}
    \begin{tabular}{cc|c|c|}
      & \multicolumn{1}{c}{} & \multicolumn{2}{c}{Vehicle $2$}\\
      & \multicolumn{1}{c}{} & \multicolumn{1}{c}{Go}  & \multicolumn{1}{c}{No Go} \\\cline{3-4}
      \multirow{2}*{Vehicle 1}  & Go & $(1,1)$ & $(2,2)$ \\\cline{3-4}
      & No Go & $(2,2)$ & $(0,0)$ \\\cline{3-4}
    \end{tabular}
\caption{One time-step model of the bottleneck reward structure. Here we make sure to reward the vehicle that exited the system, even though in an MDP its trajectory would have ended and no reward would have been given.}
\label{tab:standard_game}
\end{table}
This problem mimics the structure of the bottleneck where it is necessary to restrict the inflow to maximize the outflow. It is straightforward to see that the optimum is achieved when one vehicle goes and the other doesn't and that this is both the social optimum and a Nash equilibrium. The game is depicted in Table~\ref{tab:standard_game} where it can visually be verified that (No-Go, Go) and (Go, No-Go) are Nash Equilibria but (No Go, No Go) is not an equilibrium point. 

Open networks modify the problem in that once a vehicle exits, it ceases to receive any reward. Therefore, the vehicle that goes does not actually observe any outflow and will receive a reward of zero.
\begin{itemize}
    \item If both vehicles go, congestion occurs and they receive a reward of 1.
    \item If one vehicle goes and the other doesn't, no congestion occurs. The vehicle that went receives a reward of 0 while the other one receives a reward of 2.
    \item If both vehicles do not go, they receive a reward of 0.
\end{itemize}
\begin{table}
\centering
    \setlength{\extrarowheight}{4pt}
    \begin{tabular}{cc|c|c|}
      & \multicolumn{1}{c}{} & \multicolumn{2}{c}{Vehicle $2$}\\
      & \multicolumn{1}{c}{} & \multicolumn{1}{c}{Go}  & \multicolumn{1}{c}{No Go} \\\cline{3-4}
      \multirow{2}*{Vehicle 1}  & Go & $(1,1)$ & $(0,2)$ \\\cline{3-4}
      & No Go & $(2,0)$ & $(0,0)$ \\\cline{3-4}
    \end{tabular}
\caption{One time-step model of the bottleneck reward structure where vehicles do not receive reward after they exit.}
\label{tab:exit_game}
\end{table}
In this setting, depicted in Table.~\ref{tab:exit_game}, (No Go, No Go) is now a weak Nash Equilibrium. From the perspective of learning, this is an ubiquitous equilibrium as the vehicles that chose not to go will tend to accumulate a lot of reward. An easy solution to remove this equilibrium is to adopt the perspective of the game in Table~\ref{tab:standard_game} and continue to reward vehicles even after they leave the system. However, this would create a very noisy reward function as agents that exit the system earlier would receive a lot of reward from states and actions that they did not particularly influence. An alternative variant is to keep all the agents persistently in the system: run the system for some warm-up time to accumulate a starting number of vehicles and after that, any vehicle that exits is rerouted back to the entrance. We adopt this choice and reroute the vehicles during the training. However, this could lead to vehicles manipulating the bottleneck across reroutes and so we turn this rerouting behavior off when testing the policies after training.

\subsection{Experiment details}
\label{sec:exp_details}

We use the traffic micro-simulator SUMO~\cite{SUMO2018} for running our simulations. At training time, we use the re-routing technique discussed in Sec.~\ref{sec:nash_social} where vehicles are simply placed back at the beginning of the network after exiting. 
It is essential to note that for the multi-agent experiments we used a \emph{shared controller}, all of the agents operate in a decentralized fashion but share the same controller. 

For the training parameters for TD3, we primarily used the default parameters set in RLlib\,\cite{liang2017rllib}\footnote{\url{https://github.com/ray-project/ray/python/ray/rllib}} version 0.8.0, a distributed deep RL library. The percentage of autonomous vehicles varies among 5\%, 10\%, 20\% and 40\%. During each training rollout, we keep a fixed inflow of 2400 vehicles per hour over the whole horizon. At each time-step, a random number of vehicles are emitted from the start edge. Thus, the number of vehicles in each platoon behind the AVs will be of variable length and it is possible that at any time-step any given lane may have zero AVs in it. To populate the simulation fully with vehicles, we allow the experiment to run uncontrolled for 300 seconds as a warm-up. After that, we run an RL rollout for 1000 seconds. 

For more details, see the Appendix.

\section{Results}
\label{sec:results}
In this section we attempt to provide experimental results that answer the following questions:
\begin{enumerate}[label=H\arabic*.]
    \item \label{H1} How does the ability to improve bottleneck throughput scale with available sensing infrastructure? With penetration rate?
    \item \label{H2} Is there a single controller that will work effectively across all penetration rates?
    \item \label{H3} Can we construct an effective controller that uses purely local observations?
\end{enumerate}

\subsection{Effect of Sensing}
\label{sec:sensing_exps}

\begin{figure}[h!]
\begin{subfigure}[b]{0.48\textwidth}
     \centering
     \begin{subfigure}[b]{\textwidth}
         \centering
         \includegraphics[width=\textwidth]{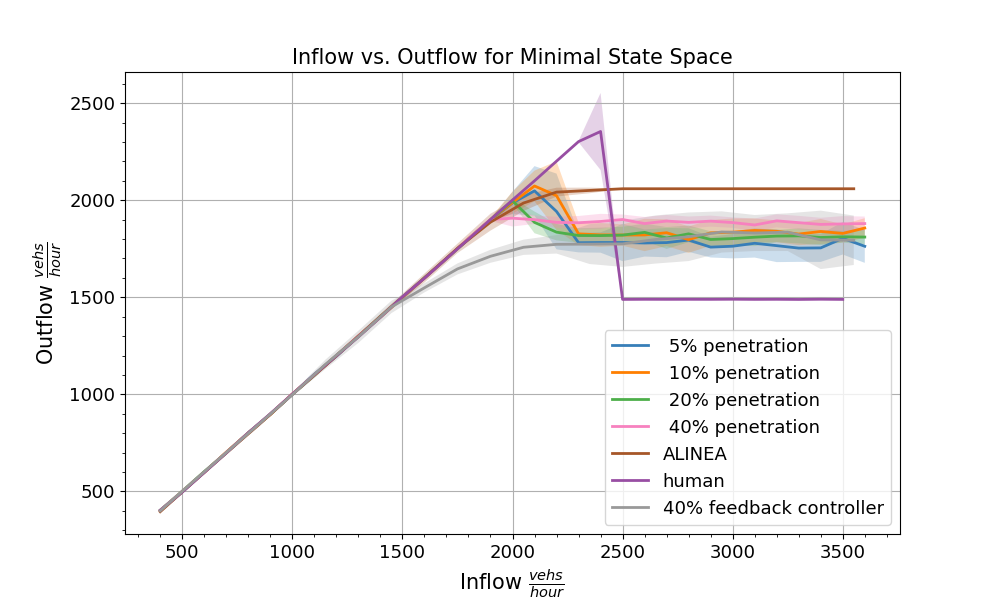}
     \end{subfigure}
     \hspace{0.05cm}
     \begin{subfigure}[b]{\textwidth}
         \centering
         \includegraphics[width=\textwidth]{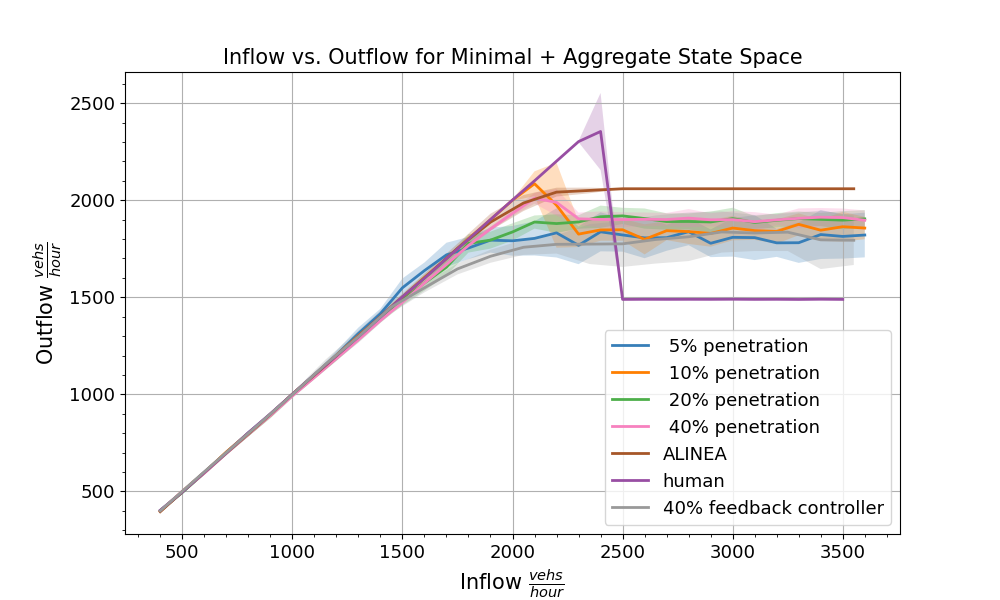}
     \end{subfigure}
          \hspace{0.05cm}
     \begin{subfigure}[b]{\textwidth}
         \centering
         \includegraphics[width=\textwidth]{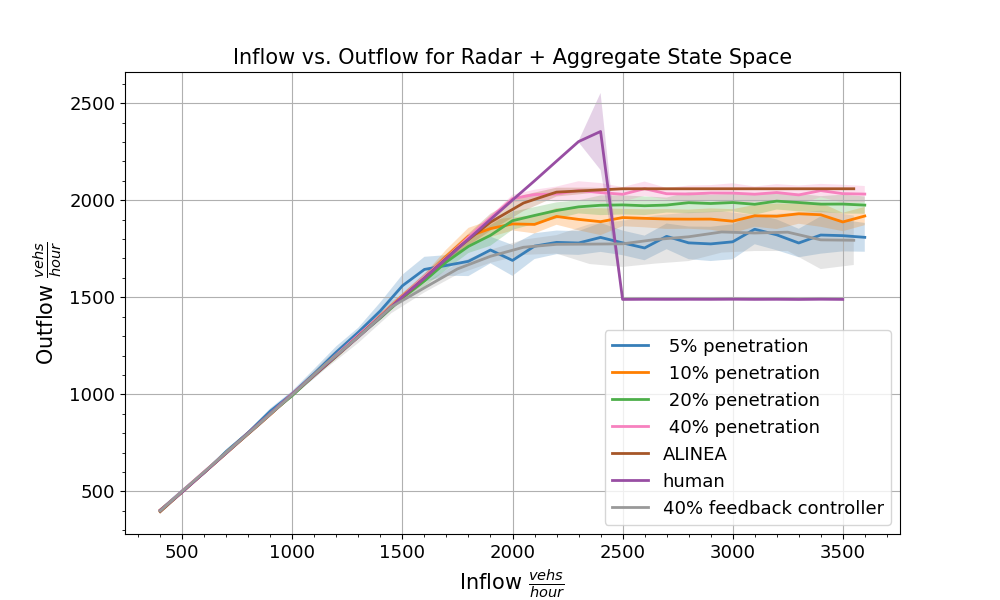}
     \end{subfigure} 
\end{subfigure}
        \caption{From top to bottom, the evolution of the inflow vs. outflow curve as penetration rates evolve for minimal, minimal + aggregate, and radar + aggregate. Within each figure we plot the performance of our controllers trained at four different penetration rates, the traffic light baseline ALINEA, the performance of our feedback controllers at a 40\% penetration rate, and the uncontrolled curve marked "human".}
        \label{fig:state_compare_agg}
\end{figure}

\begin{figure}[h!]
 \centering
 \includegraphics[width=0.48\textwidth]{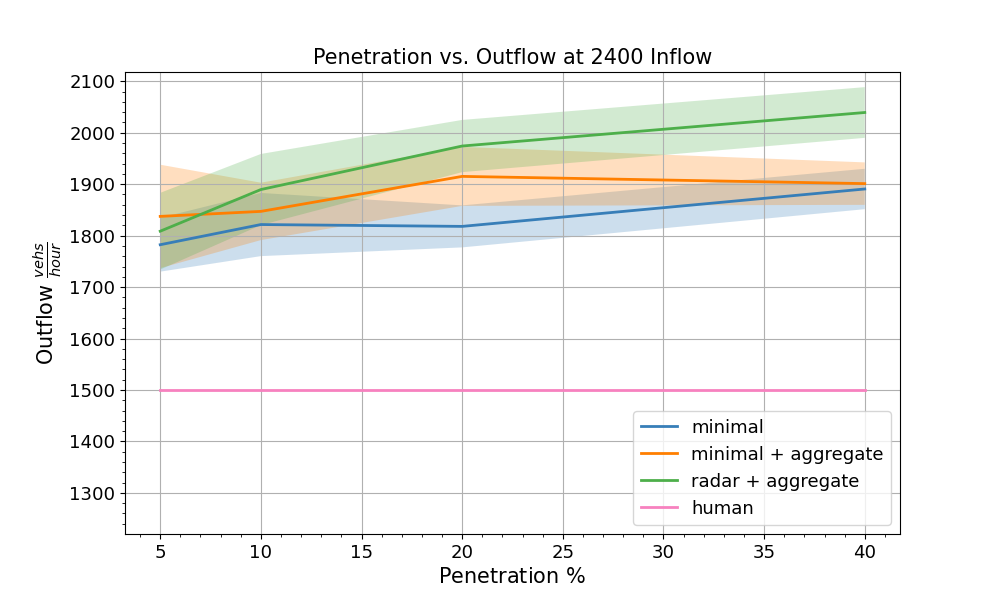}
\caption{Evolution of the outflow as a function of the penetration rate for the three state spaces we are using. We also plot the uncontrolled human baseline for reference.}
\label{fig:penetration_outflow_2400inflow}
\end{figure}

Here we compare the relative performance of the different sensing options across different penetration rates. Fig.~\ref{fig:state_compare_agg} compares the evolution of the inflow-outflow curve of the three state spaces to the uncontrolled case (labelled human), the traffic light baseline (labelled ALINEA), and the hand designed feedback controller operating at a 40\% penetration rate. Each of the state spaces outperform the hand designed feedback controller at every penetration rate and provide a 15\% improvement in the outflow even at the 5\% penetration rate.
To study the evolution of the outflow with penetration rate, Fig.~\ref{fig:penetration_outflow_2400inflow} illustrates the outflow at an inflow of 2400 as a function of penetration rate. Only the \emph{radar + aggregate} state space is able to consistently take advantage of increasing penetration rate. Excitingly, its performance at a 40\% penetration equals the performance of a traffic light based controller.
Table \ref{table:outflow_3500} summarizes the values of each of the different state spaces at an inflow of 3500 vehicles per hour.

\begin{table}[h!]
\centering
\begin{tabular}{cccc}
\toprule
 & \textbf{minimal} & \textbf{minimal + aggregate} & \textbf{radar + aggregate} \\ \midrule
5\% & 1803 $\pm$ 83 & 1813 $\pm$ 114 & 1817 $\pm$ 80 \\
10\% & 1829 $\pm$ 46 & 1863 $\pm$ 76 & 1888 $\pm$ 47 \\
20\% & 1811 $\pm$ 29 & 1897 $\pm$ 46 & 1980 $\pm$ 48 \\
40\% & 1878 $\pm$ 40 & 1910 $\pm$ 46 & 2034 $\pm$ 45 \\ \bottomrule
\end{tabular}
\caption{Average outflow and its variance at an inflow of 3500 vehicles per hour, as a function of the penetration and the state space.} 
\label{table:outflow_3500}
\end{table}


\subsection{Is There a Universal Controller?}
\label{sec:universal_controllers}

In Sec.~\ref{sec:sensing_exps} we train a separate controller for each penetration rate. Atop the additional computational expense needed to train a new controller for each data-point, having one controller per penetration rate might require accurate online estimation of current penetration rates so as to switch to the appropriate control scheme. Here we point out that at least for the controllers studied here, this concern is justified: a controller trained at one penetration rate and evaluated at another will under-perform a controller trained at the latter penetration rate. We also investigate a simple dynamics randomization strategy where we randomly sample a new penetration rate at each rollout and confirm that this can yield a controller that performs effectively across penetration rates albeit with some small loss of performance.

The key challenge is that the appropriate amount of time needed to wait before entering the bottleneck is a function of the penetration rate. As a simplified model to generate intuition, imagine that the bottleneck deterministically congests if more than 11 vehicles enter into it. We will refer to an AV with N vehicles behind it as a \emph{platoon of length N}. At a penetration rate of $10\%$, the average platoon length is 9. If we have two AV platoons ready to enter the bottleneck, one of the platoons must wait until the other platoon is almost completely into the bottleneck or else it will congest. At a $20\%$ penetration rate (platoon of length 4), however, two platoons can go at once without worrying about running into congestion.

\begin{figure}[h!]
 \centering
 \includegraphics[width=0.48\textwidth]{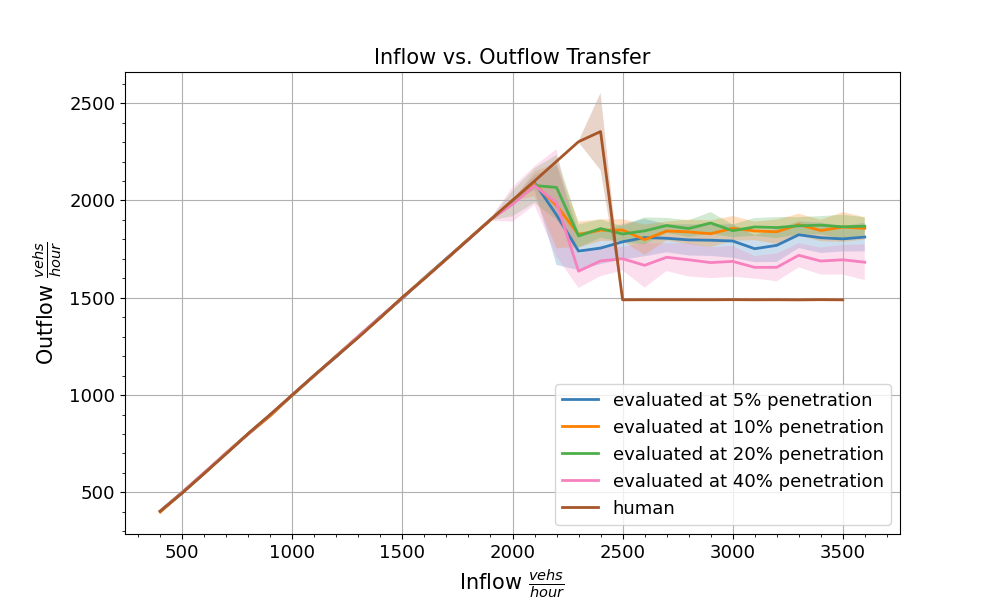}
\caption{The inflow vs. outflow curve for a controller trained at 10\% penetration on the minimal + aggregate state space, and evaluated at 5\%, 10\%, 20\% and 40\% penetration. We also plot the uncontrolled human baseline for reference.}
\label{fig:suboptimal_pen}
\end{figure}

\begin{figure}[h!]
 \centering
 \includegraphics[width=0.48\textwidth]{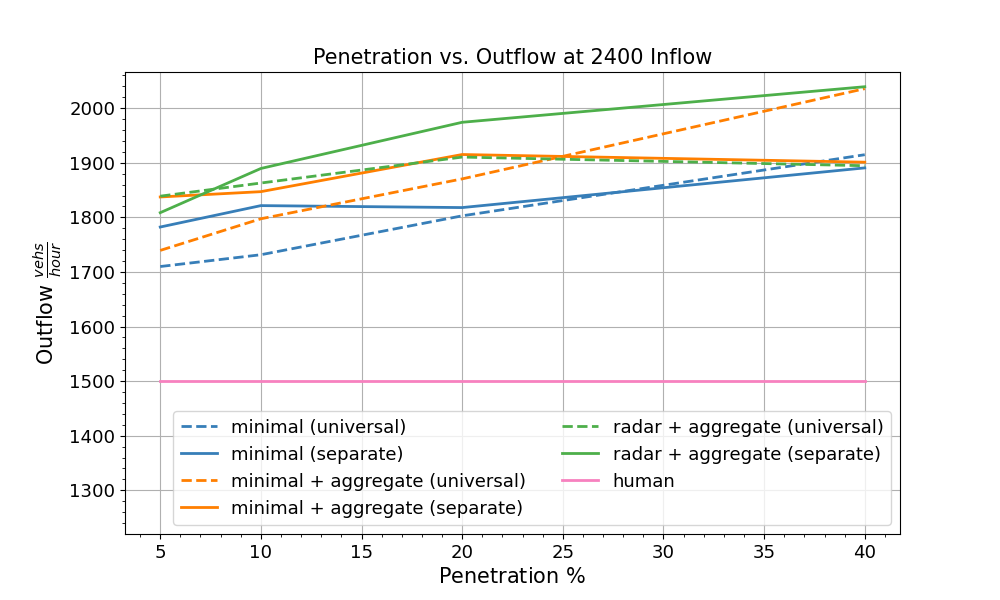}
\caption{Evolution of the outflow as a function of the penetration rate for the three state spaces we are using. For each state space, we compare the universal controller trained using dynamics randomization and evaluated at different penetration rates, to the four independent controllers trained and evaluated at their own penetrations rate of respectively 5\%, 10\%, 20\% and 40\%. We also plot the uncontrolled human baseline for reference.}
\label{fig:universal_controllers}
\end{figure}

As a result, a controller trained at low penetration rates may be too conservative when deployed at higher penetration rates while a controller trained at high penetration rates may not be conservative enough at low penetration rates. As demonstrated in Fig.~\ref{fig:suboptimal_pen}, a controller trained at a $10\%$ penetration rate does significantly worse when deployed at a 40\% penetration rate. Thus, if we use the controllers trained in Sec.~\ref{sec:sensing_exps}, it will be necessary to use either infrastructure or historical data to identify the current penetration rate and deploy the appropriate controller. This motivates our attempt to find a single controller that is stable across penetration rates.
We demonstrate that in return for some small degradation in performance, we can construct a single controller that performs robustly across penetration rates. To achieve this, we use \emph{dynamics randomization} and for each trajectory we sample a new penetration rate $p$ uniformly from $p\sim U(0.05, 0.4)$. We refer to these as \emph{universal controllers} and the controllers trained at individual penetration rates in Sec.~\ref{sec:sensing_exps} as \emph{independent} controllers. Fig.~\ref{fig:universal_controllers} shows the performance of the universal controllers compared to the controllers trained at a penetration rate and evaluated at the same penetration rate, for each of the three state spaces we are using. 

From Fig.~\ref{fig:universal_controllers}, we can see that the results do not have a consistent trend as the universal controllers both over-perform and under-perform at different penetration rates. For example, the minimal + aggregate controller gives up 100 vehicles per hour at a 5\% penetration rate, but outperforms the independent controller by 250 vehicles per hour at high penetration rates. However, we note that even the worst outcome at low penetration rates, the universal minimal controller, outperforms the uncontrolled baseline of 1550 vehicles per hour. Additionally, the universal radar + aggregate controller consistently provides an outflow of 1850 vehicle per hour which achieves the desired goal of an effective controller independent of penetration rate.



\subsection{Controllers without macroscopic observations}
\label{sec:nocongestnb}

The three state spaces studied earlier, minimal, minimal + aggregate, radar + aggregate, all contain the number of vehicles in the bottleneck (edge 4 in Fig.~\ref{fig:segments}) in the state space as well as average speed data on edges 3, 4, 5 in the aggregate cases. Acquiring this information consistently (rather than through a lucky LIDAR or radar bounce picking up many vehicles ahead of the ego vehicle) would require either camera or loop sensing infrastructure. We would like to understand whether efficient control can be done without access to the number of vehicles in the bottleneck, a quantity we refer to as \emph{congest number}. This would enable us to perform control that is fully decentralized in both action and observation, allowing us to deploy these systems with no additional infrastructure cost. 

We attempt to answer this question by using the radar state space alone without any aggregate information. Thus, the vehicle only has access to the speeds and distances to the vehicles directly ahead of and behind it in each lane.  Although it is not obvious how the controller will accomplish inference of the \emph{congest number}, a few possible options:

\begin{enumerate}
    \item There exists a scheme that does not actually depend on the number of vehicles in the bottleneck. 
    \item The number of vehicles in the bottleneck can be inferred from the distance to the visible vehicles or the speed of the visible vehicles. For example, if a vehicle observed in the bottleneck is stopped that likely indicates congestion while high velocities would indicate free flow.
    \item Vehicles can learn to communicate through motion by adopting vehicle spacing and velocities that observing vehicles can use to infer the \emph{congest number}. For example, a velocity between 2 and 3 m/s could indicate 0-5 vehicles in the bottleneck, between 3 and 4 m/s could indicate 6-10 vehicles and so on.
\end{enumerate}

\begin{figure}[h!]
 \centering
 \includegraphics[width=0.48\textwidth]{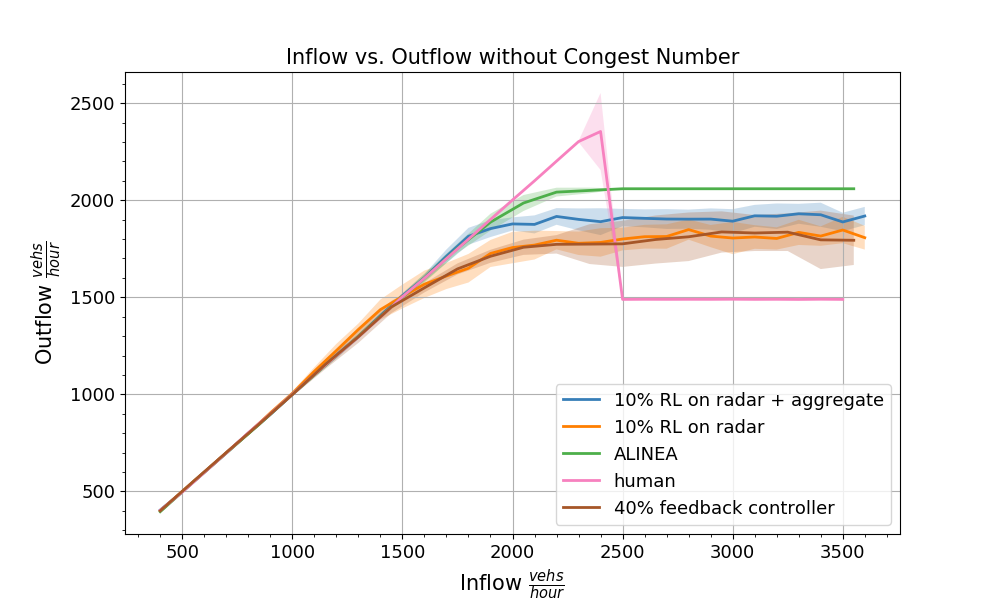}
\caption{Evolution of the inflow vs. outflow curve for controllers trained at a penetration rate of 10\%. We compare a controller trained on the full radar + aggregate state space to a controller only trained on the radar state space, which means it doesn't have access to the number of vehicles in the bottleneck. We also plot the performance of the traffic light baseline ALINEA, of our feedback controller at 40\%, and of the uncontrolled curve marked "human".}
\label{fig:no_congest_number_training}
\end{figure}

While we are not able to conclusively establish which of the above hypotheses are active, Fig.~\ref{fig:no_congest_number_training} demonstrates the effect of removing any macroscopic information from the state space. The radar state space with no macroscopic data, marked in orange, deviates from the aggregate state space by about 100 vehicles per hour but still sharply outperforms an uncontrolled baseline (marked human). Since radar sensors are now standard features on level 2 vehicles, this suggests that a fully decentralized controller can be deployed using available technology. 

\subsection{Robustness of Simplifications}
\label{sec:ablations}

In this section we attempt to relax some of the simplifying assumptions made in the design of our problem. Specifically, we investigate the effects of the following changes:
\begin{itemize}
    \item Lane changing. In the prior results, we have disabled lane changing. We now study whether our best controllers are able to handle adding lane changing to the human driver dynamics without retraining the RL controller.
    \item Simplifications in the "radar" model. Our "radar" state space does not take into account occlusions and can return a vehicle an arbitrary distance away. 
\end{itemize}

\begin{figure}[h!]
 \centering
 \includegraphics[width=0.48\textwidth]{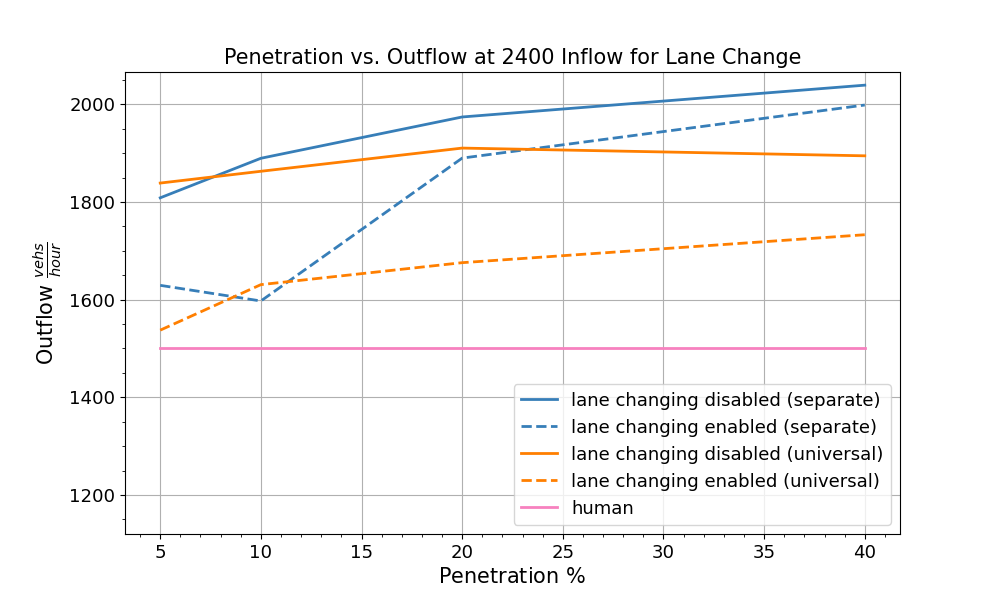}
\caption{Evolution of the outflow as a function of the penetration rate for controllers trained on the radar + aggregate state space. We compare controllers that have been trained with lane changing disabled, to those sames controllers when lane changing is enabled at evaluation time. We compare both controllers trained on a fixed penetration rate of 5\%, 10\%, 20\% or 40\%, referred to as "separate", and controllers trained at a random penetration rate between 5\% and 40\% as explained in \ref{sec:universal_controllers}, referred to as "universal". We also plot the uncontrolled human baseline for reference.}
\label{fig:lane_change_eval}
\end{figure}

In Fig.~\ref{fig:lane_change_eval}, we enable lane changing and examine how effective our controllers are as the penetration rate evolves. Unsurprisingly, at low penetration rates there is a sharp reduction in outflow relative to the lane changing disabled setting. The challenge is that when one of the AVs goes, other vehicles will rapidly lane change into its lane which prevents the AV from restricting the inflow. As the penetration rate increases, when an AV at the front of the queue goes, a new AV rapidly arrives to replace it which consequently minimizes the impact of lane changing.
However, we note that when vehicles lane change in SUMO, they instantly change lanes which may enable more aggressive lane changing than is physically possible. Hence, the degradation in outflow might be lower in reality than it is in our simulator. Furthermore, disabling lane changing at a bottleneck by painting new road lines should be relatively cheap.

\begin{figure}[h!]
 \centering
 \includegraphics[width=0.48\textwidth]{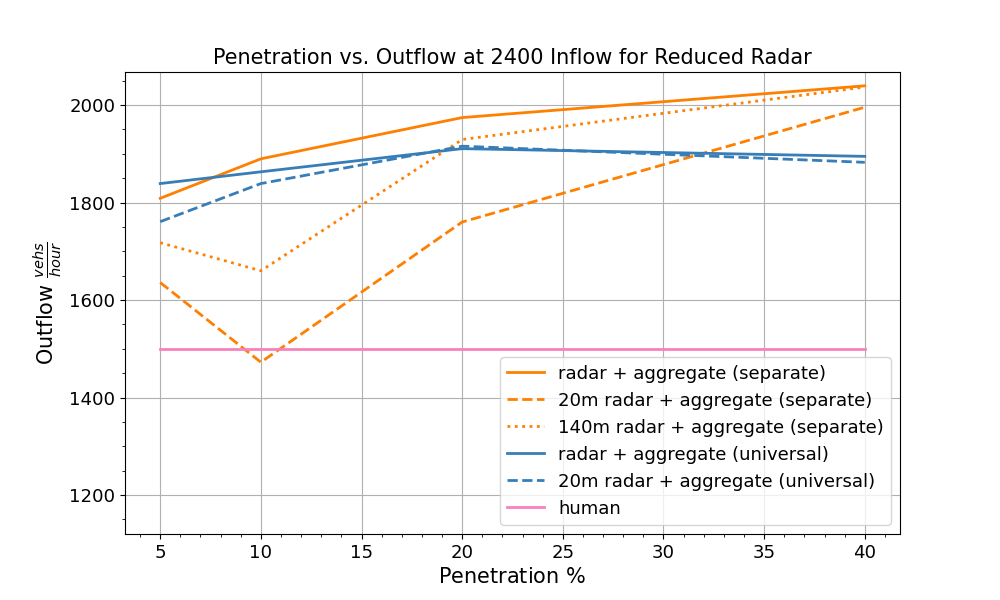}
\caption{Evolution of the outflow as a function of the penetration rate for controllers trained on the radar + aggregate state space. We compare controllers that have been trained with the (normal) entire radar state space, to those same controllers when the radar is restricted to seeing only vehicles up to a distance of 20 meters or 140 meters at evaluation time. We compare both controllers trained on a fixed penetration rate of 5\%, 10\%, 20\% or 40\%, referred to as "separate", and controllers trained at a random penetration rate between 5\% and 40\% as explained in \ref{sec:universal_controllers}, referred to as "universal". We also plot the uncontrolled human baseline for reference.}
\label{fig:reduced_radar_eval}
\end{figure}

As for the question of a more "realistic" radar model, Fig.~\ref{fig:reduced_radar_eval} presents our attempt to restrict the range beyond which vehicles cease to become visible. We take a trained policy and cap how far it is allowed to see as an approximation of a restricted radar range. We try two restrictions, not seeing past 20 meters and not seeing past 140 meters. If there is no vehicle within that range, we set a default state with a distance of $20$ or $140$ meters respectively, a speed of $5$ meters per second, and treat it as a human vehicle by passing a zero to the boolean that indicates whether a vehicle is human or autonomous. Other replacements in the state for vehicles that are too far away to see are possible, we could instead replace the missing states with a vector of $-1$'s but found empirically that the replacement discussed led to better performance. We find that the \emph{universal} controllers are relatively robust to this replacement which suggests that their controller is more independent of actual vehicle sensing and more dependent on macroscopic states. This ablation, simply ignoring vehicles that are too far away, is an extremely approximate model of how radar might work and replacing it with more accurate radar models is a topic for future work.

\subsection{Instability of reward curve}
\label{sec:instability}
For purposes of reproducibility, we provide a few representative reward curves from some of the training runs. These should help establish a sense of what the expected reward is as well as provide a calibrated sense of what fraction of training runs are expected to succeed. Since we use 36 CPU machines and each training run requires 1 CPU, we are able to train 35 random seeds in parallel (1 CPU is used to manage all the trainings) and we keep the best performing one. The following figure presents the results of 9 of these random seed trials (for better visibility) from one of the training runs.

\begin{figure}[h!]
 \centering
 \includegraphics[width=0.48\textwidth]{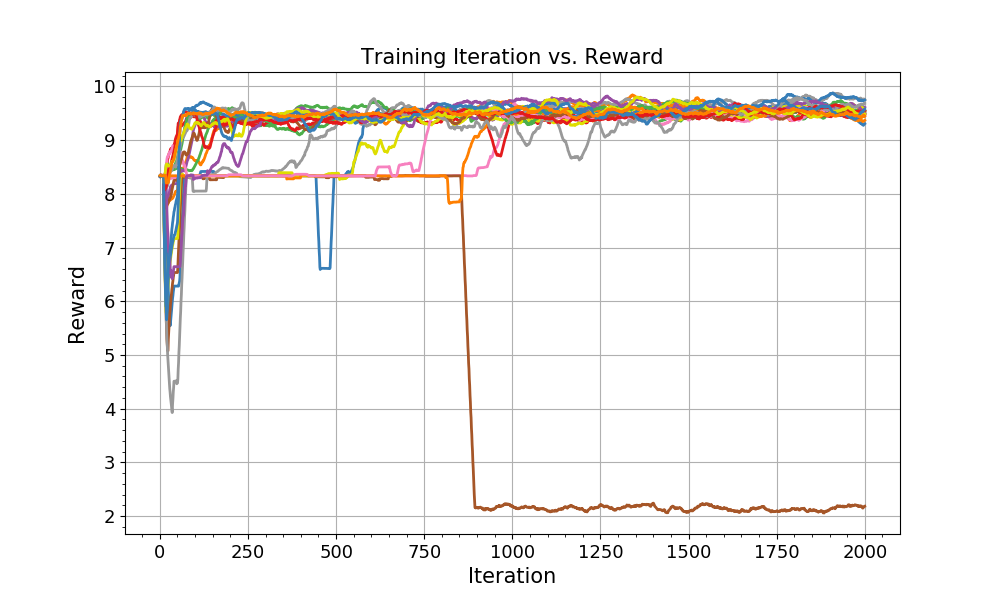}
\caption{A sample of runs from a seed sweep at a penetration rate of 10\%. Most of the seeds converge to a good policy.}
\label{fig:stable_rewards}
\end{figure}

\begin{figure}[h!]
 \centering
 \includegraphics[width=0.48\textwidth]{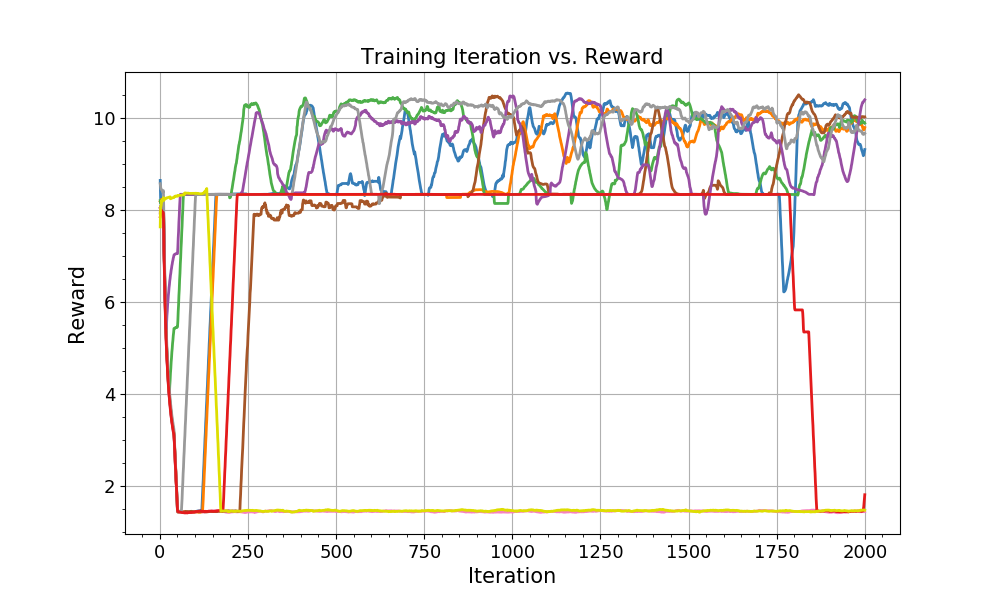}
\caption{A sample of runs from a seed sweep at a penetration rate of 40\%. There is significant instability that is expected from using single-agent algorithms in a multi-agent problem.}
\label{fig:unstable_rewards}
\end{figure}

Fig.~\ref{fig:stable_rewards} and Fig.~\ref{fig:unstable_rewards} represent reward curves from low penetration rate runs (10\%) and high penetration rate runs (40\%) respectively. A high scoring run corresponds to an average agent reward around 10, a reward slightly above eight corresponds to all the vehicles just zooming into the bottleneck without pausing, and a reward below 2 corresponds to the vehicles mostly coming to a full stop. As is clear from the figures, at low penetration rates the training is relatively stable and converges quickly. However, as the penetration rate increases the reward curves become extremely unstable, with rapid oscillations in the expected reward. This instability is not due to variations in the outflow as the std. deviation of the outflow is low but is likely the outcome of applying independent Q-learning in a multi-agent system, leading to non-stationarity in the environment. Methods that explicitly handle this non-stationarity by using a centralized critic such as MADDPG~\cite{lowe2017multi} may help reduce the instability in the training.

\section{Conclusions and Future work}
\label{sec:conclusions}
In this work we demonstrated that low levels of autonomous penetration, in this case 5\%, are sufficient to learn an effective flow regulation strategy for a severe bottleneck. We demonstrate that even at low autonomous vehicle penetration rates, the controller is able to improve the outflow by 300 vehicles per hour. Furthermore, we are able to use additional availability of AVs and at a 40\% penetration rate get equal performance to actually installing a new traffic light to regulate the inflow. 

However, many open questions remain. In this work we use an independent Q-learning algorithm which leads to serious instability in the training. As discussed in Sec.~\ref{sec:instability}, at high penetration rates many of the training runs become unstable, which makes it unclear if we are near to the optimal policy for those penetration rates. This is a challenging multi-agent RL task and it would be interesting to see whether multi-agent RL algorithms that use a centralized critic like MADDPG~\cite{lowe2017multi} and QMIX~\cite{rashid2018qmix} would lead to more stable training. Furthermore, the training procedure takes 24 hours so finding algorithms that can perform with higher sample efficiency is critical.

One possible direction to pursue in future work is to increase the level of realism, adding both lane changing and an accurate radar model that correctly accounts for obscurity. In preliminary investigations in Sec.~\ref{sec:ablations}, we found that lane changing degrades the performance of our controllers as cars simply lane change into the lane that is currently moving and thus avoid inflow restriction. It may be the case that this behavior can be avoided by more complex coordination between the AVs in which they explicitly arrange themselves to block this lane changing behavior. Preliminary experiments we ran in which training was done with lane changing on did not yield particularly strong results but this may be an artifact of our choice of training algorithm.

Another open question is to investigate the effects of coordination between the vehicles. In the decentralized case, we still do not know the extent to which the AVs are coordinating in their choice of action. While implicit coordination is possible due to vehicles being aware of which nearby vehicles are also autonomous, we have only provided circumstantial evidence that this is actually occurring. In the case of lane changing, such coordination may be needed to prevent human drivers from skipping lanes and decreasing the total outflow. Additionally, we do not use memory in any of our models which may be limiting the effectiveness of our controllers. Using memory-based networks such as LSTMs could be an interesting direction of future work.

One approach we intend to explore to explicitly enable coordination is to allow the AVs to communicate amongst themselves. Perhaps by broadcasting signals to nearby vehicles, the AVs can learn to coordinate platoons in such a way that changing lanes no longer appears advantageous to the human driver and they will remain in their lane. Furthermore, if communication proves useful, it is possible that the AVs may develop a "language" that they use for coordination. Examining whether such a "language" emerges is a future thread of work.

Finally, an exciting potential consequence of these results, given that they're decentralized and only use local information available via vision, is that human drivers could potentially implement these behaviors. Investigating whether this scheme could be deployed via human driving, whether by constructing a mobile app that provides instructions or by teaching a new driving behavior, is a direction we hope to explore in the future. 

\bibliographystyle{ieeetr}
\bibliography{bottleneck_bib}

\section{Acknowledgements*}
Eugene Vinitsky is a recipient of an NSF Graduate Research Fellowship and funded by the National Science Foundation under Grant Number CNS-1837244. Computational resources for this work were provided by an AWS Machine Learning Research grant. This material is also based upon work supported by the U.S. Department of Energy’s Office of Energy Efficiency and Renewable Energy (EERE) award number CID DE-EE0008872. The views expressed herein do not necessarily represent the views of the U.S. Department of Energy or the United States Government.


\clearpage
\appendix

\subsection{Detailed MDP}
\label{sec:detailed_mdp}
Here we provide significantly more details about the MDP defined in Sec.~\ref{sec:mdp_construction}. We have three possible state spaces that we investigate.
The first state set we call the \emph{radar state} as the states accessed would be readily available via an on-board radar and GPS. This is the state space that would be most easily implemented using existing technology on an autonomous vehicle. In the radar environment, the state set is:
\begin{itemize}
    \item The speed and headway of one vehicle ahead in each of the lanes and one vehicle behind. If the vehicle is on a segment with four lanes, it will see one vehicle ahead in each of the lanes and one vehicle behind in each of the lanes. A missing vehicle is indicated as two zeros in the appropriate position. Subject to some restrictions on sensing range, this information can be acquired via radar. Refer to Fig. \ref{fig:state_spaces} for a diagrammatic description.
    \item The speed of the ego vehicle as well as its lane where lanes are numbered in increasing order from right to left.
    \item The edge number and position on the edge of the ego vehicle where the edge numbering is according to Fig.~\ref{fig:segments}. This information would readily be available via GPS. We supplement this with an "absolute" position on the network, indicating how far the vehicle has traveled. This latter state is technically redundant and could be inferred from edge number and position.
    \item A counter that indicates how long the speed of the vehicle has been below $0.2$ meters per second. This is used to endow the AV with a memory that allows it to track how long it has been stopped / waiting at the bottleneck entrance.
    \item A global time counter indicating how much time has passed.
\end{itemize}
Note that in the \emph{radar state} that information about the bottleneck is only indirectly available; it can only examine the states of visible vehicles and use it to infer information about the bottleneck state.

The second set of states would be available given appropriate loop sensing infrastructure or a sufficient number of overhead cameras distributed throughout the bottleneck. These states endow the AV with macroscopic information about the bottleneck. We refer to this as the \emph{aggregate state}. Here the additional states are:
\begin{itemize}
    \item The average speed of the vehicles on edges 3, 4, and 5.
    \item The number of vehicles in the bottleneck.
    \item A global time counter indicating how much time has passed.
\end{itemize}

Finally, the final set of states we consider is a significantly pruned state set in which we have hand-picked what we believe to be a minimal set of states with which the task can be accomplished. We refer to this as the \emph{minimal state}. This state space should yield the fastest learning due to its small size. Here the states are:
\begin{itemize}
    \item Total distance travelled.
    \item The number of vehicles in the bottleneck.
    \item A counter that indicates how long the speed of the vehicle has been below $0.2$ meters per second.
    \item Ego speed.
    \item Leader speed.
    \item Headway.
    \item The amount of time our feedback controller described in Sec.~\ref{sec:feedback_control} would wait before entering the bottleneck. This state is intended to ease the learning process since initially the vehicles can simply learn to imitate this value.
\end{itemize}

From these three potential sets of states we form three combined state spaces that we study: radar + aggregate, minimal + aggregate, and minimal alone. Each of these represents a different set of assumptions on what sensing technology will be available, as is illustrated in Fig. \ref{fig:state_spaces}. We characterize the relative performance of these different state spaces in Sec.~\ref{sec:sensing_exps}.

The action space is simply a 1-dimensional acceleration. The vehicles enter the network at $25$ meters per second and roughly maintain that speed as they travel along. 
Since our intent is for the controllers to stop at edge three and determine the optimal time to enter the bottleneck, we want to increase the likelihood of them coming to a stop on edge three. To increase the likelihood of sharp decelerations, we bound our action space between $\frac{1}{8} \left[-4.5, 2.6\right]$ and re-scale the actions by multiplying them by eight. The neural network weight initialization scheme we use (see appendix for details) tend to output actions bounded between $\left[-1, 1\right]$ at initialization time; this re-scaling scheme makes it likelier that large decelerations are applied.

While we could include lane changes as a possible action, we made the assumption that it was unlikely that lane-changing behavior could be a positive and would only cause the training to take longer. To prevent the vehicles from forming unusual patterns at the entrance of the bottleneck, control is only applied on edge 3 (edges numbered according to Fig.~\ref{fig:segments}). However, states and rewards are received at every time-step and consequently actions are computed at each time-step: we simply ignore the controller output unless we are on the third edge.

We are trying to optimize throughput, so as our reward function we simply use the number of vehicles that have exited in the previous time-step as a reward
\begin{equation*}
    r_t(s_t, a_t) = n_t / N
\end{equation*}
where $n_t$ is the number of vehicles that have exited in that time-step and $N$ is a normalization term that was use to keep the cumulative reward at reasonable values. We use $N=50$ in this work.
Since the outflow is exactly the quantity we are trying to optimize, optimizing our global reward function should result in the desired improvement. This is a global reward function that is shared by every agent in the network. 

However, we note a few challenges that make this a difficult reward function optimize. First, the reward is global which causes difficulties in credit assignment. Namely, it is not clear which vehicle's action contributed to the reward at any given time-step. Secondly, there is a large gap between when an action is taken and when the reward is received for that action. That is, a vehicle choosing to enter the bottleneck does not receive any reward directly attributable to that decision for upwards of 20 steps. Finally, the bottleneck being fully congested is likely a local minimum that is hard to escape. Once congestion has onset, it cannot be removed without a temporary period where the inflow into the bottleneck is reduced. However, a single vehicle choosing to not enter the bottleneck would have negligible effect on the inflow, making it difficult for vehicles to learn that decongestion is even possible.

\subsection{Training parameters}
Here we outline the optimal hyperparameters and seed for every experiment presented in this paper. These hyperparameters were found, as discussed in Sec.~\ref{sec:exp_details}, by sweeping a fixed set of hyperparameters, picking the policy with the highest reward after 2000 iterations, and then sweeping 35 seeds.

\begin{table*}[h!]
\centering
\begin{tabular}{@{}lllllll@{}}
\toprule
            &           & \texttt{actor\_lr} & \texttt{critic\_lr} & \texttt{n\_step} & \texttt{prioritzed\_replay} & \texttt{seed} \\ \midrule
            & 5\%       & 0.001    & 0.0001    & 5      & True               & 24   \\
            & 10\%      & 0.0001   & 0.0001    & 5      & False              & 29   \\
minimal     & 20\%      & 0.0001   & 0.001     & 5      & True               & 15   \\
            & 40\%      & 0.0001   & 0.0001    & 5      & False              & 9    \\
            & universal & 0.0001   & 0.0001    & 5      & False              & None \\  \midrule
            & 5\%       & 0.0001   & 0.0001    & 5      & False              & 28   \\
            & 10\%      & 0.001    & 0.0001    & 5      & True               & 3    \\
minimal     & 20\%      & 0.0001   & 0.0001    & 5      & False              & 0 \\
+ aggregate & 40\%      & 0.0001   & 0.001     & 5      & True               & 9    \\
            & universal & 0.001    & 0.001     & 5      & False              & None \\  \midrule
            & 5\%       & 0.0001   & 0.0001    & 5      & True               & 14   \\
            & 10\%      & 0.0001   & 0.001     & 5      & False              & 16   \\
radar       & 20\%      & 0.0001   & 0.001     & 5      & False              & 17   \\
+ aggregate & 40\%      & 0.0001   & 0.001     & 5      & True               & 19   \\
            & universal & 0.0001   & 0.0001    & 5      & False              & None \\ \midrule
no congest number & 5\%       & 0.0001   & 0.001    & 5      & False               & None   \\ \bottomrule
\end{tabular}
\caption{Parameters used during training with RLlib~\cite{liang2018rllib}'s implementation of the TD3 algorithm for the experiments trained at penetrations of 5\%, 10\%, 20\%, 40\% or universally (cf. Sec. \ref{sec:universal_controllers}) for the minimal, minimal + aggregate and radar + aggregate state spaces. The experiments trained at fixed penetration have been trained for 2000 iterations with a parameter search (cf. Sec \ref{sec:exp_details}) followed by a grid search on the best parameters (cf. Sec \ref{sec:reproducibility}), after which the best seed was kept. The universal controllers have been trained for 400, 1200 and 2000 iterations for respectively the minimal, minimal + aggregate and radar + aggregate state spaces, and no seed search was ran for these three experiments. The last line of the table refers to the experiment that was trained without macroscopic information about the bottleneck's outflow (cf. Sec \ref{sec:nocongestnb}); it was trained for 1600 iterations and without seed search. Only parameters that differ from the default RLlib configuration for TD3 (\url{https://docs.ray.io/en/releases-0.6.6/rllib-algorithms.html\#deep-deterministic-policy-gradients-ddpg-td3}) are detailed here.}
\end{table*}

\subsection{Experiment details}
\label{sec:exp_details}

For the training parameters for TD3, we primarily used the default parameters set in RLlib\,\cite{liang2017rllib}\footnote{\url{https://github.com/ray-project/ray/python/ray/rllib}} version 0.8.0, a distributed deep RL library. Both the policy and the Q-function are approximated by a neural network, each with two hidden layers of size $[400, 300]$ and a ReLU non-linearity following each hidden layer. We used a training buffer size of $100000$ samples and use a ratio of $5$ new samples from the environment for every gradient step.
For each training run, we also perform a hyperparameter sweep over the following values:
\begin{itemize}
    \item The learning rate for both the policy and the critic: $\left[1e-3, 1e-4\right]$.
    \item The length of the reward sequence before truncating the target with the Q-function (also known as \emph{n-step return}): $\left[1, 10\right]$
    \item We test both using and not using prioritized experience replay~\cite{schaul2015prioritized}.
\end{itemize}
The best performing value, in terms of final converged reward, is selected from the hyperparameters, after what we run 35 random seeds using the best hyperparameters. We select the highest reward at the end of training from these random seeds.



The percentage of autonomous vehicles varies among 5\%, 10\%, 20\% and 40\%. During each training rollout, we keep a fixed inflow of 2400 vehicles per hour over the whole horizon. At each time-step, a random number of vehicles are emitted from the start edge. Thus, the number of vehicles in each platoon behind the AVs will be of variable length and it is possible that at any time-step any given lane may have zero AVs in it. To populate the simulation fully with vehicles, we allow the experiment to run uncontrolled for 300 seconds. After that, the horizon is set to 1000 more seconds. 

At training time, we use the re-routing technique discussed in Sec.~\ref{sec:nash_social} where vehicles are simply placed back at the beginning of the network after exiting. However, when performing the inflow-outflow sweeps to evaluate the efficacy of the policy / generate the graphs in this paper, we turn rerouting off to ensure that our policy's performance is not dependent on the policy using the rerouting to generate some unusual behavior. To compute the outflow at a given inflow value, we run the system for 1000 seconds and compute the outflow over the last 500 seconds.

We use the traffic micro-simulator SUMO~\cite{SUMO2018} for running our simulations. We use a simulation step of 0.5 seconds and a first order Euler integration for the dynamics. While we use a relatively small time-step to maintain sensible dynamics, we use action repetition and only select a new controller action every 2.5 seconds. Each action is thus repeated five times; this approach is useful for speeding up training when the system dynamics change at a slower time-scale than the dynamics update frequency. This technique is standard when applying deep RL to Atari games~\cite{mnih2013playing}. Similarly, states, rewards and actions are only computed once every 5 simulation steps during both training and evaluation. Thus, running the system for 1000 seconds corresponds to 400 environment steps but to 2000 simulation steps.

It is essential to note that for the multi-agent experiments we used a \emph{shared controller}, all of the agents operate in a decentralized fashion but share the same controller. 

\subsection{Reproducibility and Experimental Details}
\label{sec:reproducibility}

The code used to run the experiments and plot all of the figures is available at our fork of \texttt{Flow}\footnote{\url{https://github.com/eugenevinitsky/decentralized_bottlenecks}}.

For each example, we perform the hyperparameter sweep indicated in Sec. \ref{sec:exp_details} and train at a fixed inflow of 2400 vehicles per hour. We then take the best hyperparameter set and re-run the experiment using 35 different seeds. The seed with the highest outflow is taken as the controller for each example. 

All experiments are run on c4.8xlarge machines on AWS EC2 which have 36 virtual cores each. Since we use a single-processor implementation of TD3, both our hyperparameter sweeps and seed sweeps fit on a single machine. 


\subsection{Feedback Controller Sweep Parameters}
For our feedback controllers, we swept the following hyperparameters in a grid:
\begin{itemize}
    \item $n_\text{crit}$ = [6, 8, 10]
    \item $K$ = [1, 5, 10, 20, 50]
    \item $q_\text{init}$ = [200, 600, 1000, 5000, 10000]
\end{itemize}
Empirically, we found that these were the parameters that the control scheme was most sensitive to. Whenever a vehicle enters, we set $q_0 = q_\text{init}$ so each vehicle is maintaining its own counter of the appropriate wait-time.

\end{document}